\documentclass[prd,preprintnumbers,nofootinbib,tightenlines,superscriptaddress]{revtex4-2}
\usepackage{graphicx,epstopdf}
\usepackage{amsmath,amssymb,graphicx}
\usepackage[usenames, dvipsnames]{color}
\usepackage{slashed}
\usepackage{times}
\usepackage{latexsym}
\usepackage{longtable}
\usepackage[utf8]{inputenc}
\usepackage{hyperref}

\newcommand{\orcid}[1]{\href{https://orcid.org/#1}{\includegraphics[width=8pt]{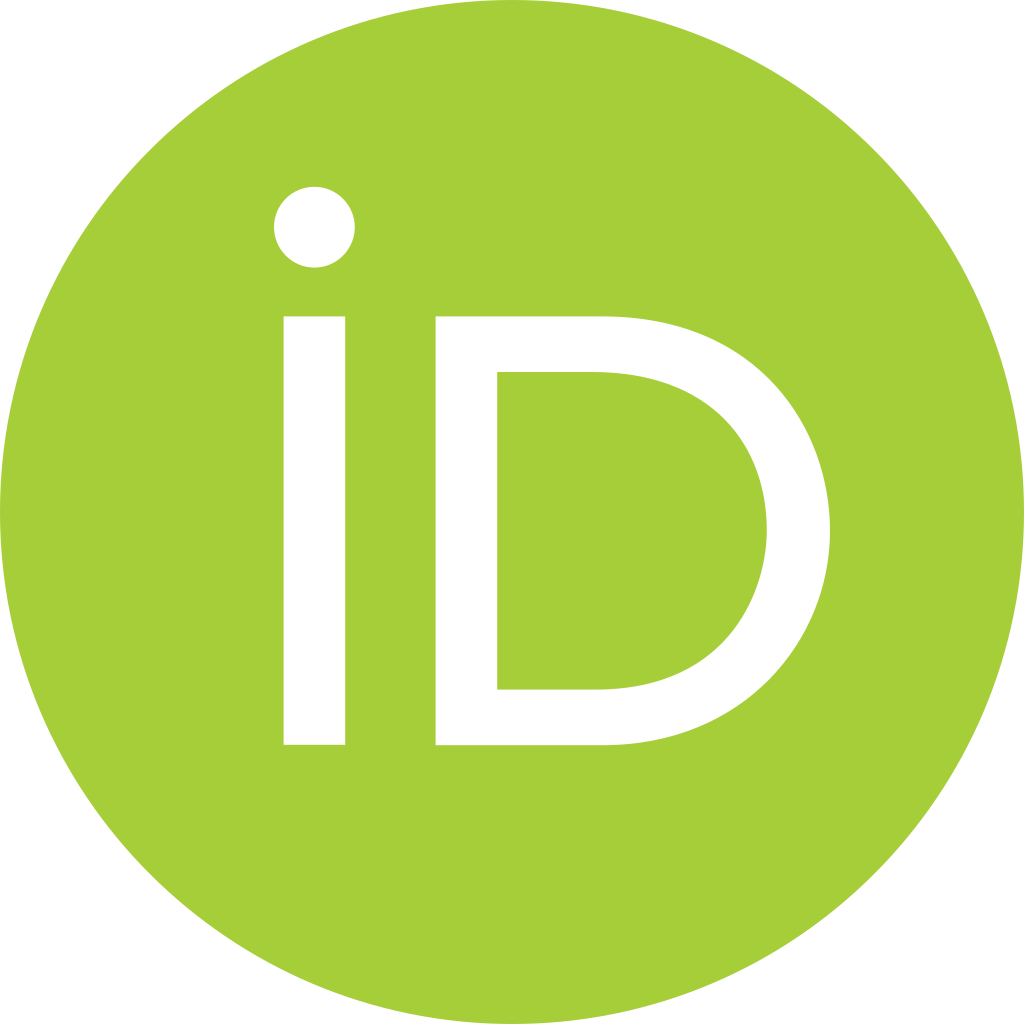}}}

\begin{document}
\title{Relaxing Constraints on Dark Matter Annihilation Near the Supermassive Black Hole in M87}

\author{Mehrdad Phoroutan-Mehr \orcid{0000-0002-9561-0965}}
\email[]{mphor001@ucr.edu}
\affiliation{Department of Physics \& Astronomy, University of California, Riverside, CA 92521, USA }

\author{Hai-Bo Yu \orcid{0000-0002-8421-8597}}
\email[]{haiboyu@ucr.edu}
\affiliation{Department of Physics \& Astronomy, University of California, Riverside, CA 92521, USA }

\date{\today}

\begin{abstract}

The supermassive black hole at the center of M87 could redistribute dark matter particles within its sphere of influence, creating a high-density region known as a density spike. This spike can significantly enhance dark matter annihilation signals, making M87 a critical target for deriving stringent constraints on annihilation cross sections. In this work, we demonstrate that these constraints are highly sensitive to the choice of the halo density profile for M87. Motivated by recent kinematic studies of M87, we adopt a cored halo model and find that the constraints on dark matter annihilation are significantly relaxed. Specifically, in the cored halo scenario, the smooth part of the halo overwhelmingly dominates the annihilation signals, whereas the commonly-assumed cuspy halo model attributes a major contribution to the spike. We demonstrate this effect using a dark matter model with a light mediator.

\end{abstract}

\maketitle

\section{Introduction}

It is well established that dark matter makes up about $85\%$ of the mass content of the Universe~\cite{Planck:2015fie}, but its particle nature remains elusive. The weakly interacting massive particle (WIMP) is a well-studied dark matter candidate as it could lead to signals in terrestrial dark matter searches; see~\cite{Roszkowski:2017nbc,Cirelli:2024ssz}. For example, WIMPs could annihilate to standard model particles, such as the electron and photon, which could be detected in indirect detection experiments. Since the annihilation rate increases with the dark matter density in the halo, the central region of galaxies is an important target area. Dark matter could form a density spike surrounding a black hole at the center of galaxies~\cite{Gondolo:1999ef}, further boosting annihilation signals. In fact, observations of the giant elliptical galaxy M87, which hosts a supermassive black hole of $6.5\times10^9~{\rm M_\odot}$~\cite{EventHorizonTelescope:2019dse}, put strong constraints on WIMP models~\cite{Lacroix:2015lxa,Lacroix:2016qpq,Yuan:2021mzi,Chen:2024nua}. These studies assume that the M87 halo follows a Navarro-Frenk-White (NFW) density profile~\cite{Navarro:1996gj} and the density spike near its supermassive black hole (SMBH) has a profile of $\rho_{\rm sp}\propto r^{-7/3}$. The signal strength from the spike could be larger than that from the smooth part of the halo, and the constraints from M87 could rule out thermal WIMPs. Nevertheless, Ref.~\cite{Alvarez:2020fyo} shows that such constraints could be significantly weakened in self-interacting dark matter (SIDM) models~\cite{Tulin:2017ara}, which predict a shallower profile for the spike $\rho_{\rm sp}\propto r^{(-3+a)/4}$~\cite{Shapiro:2014oha}, where $a$ characterizes the velocity dependence of the self-scattering cross section $\sigma\propto v^{-a}$. Even in the case of Coulomb-like self-interactions with $a=4$, the SIDM spike has $\rho_{\rm sp}\propto r^{-7/4}$, weaker than the spike $\rho_{\rm sp}\propto r^{-7/3}$ for WIMPs. 

Interestingly, recent measurements of M87 based on multiple kinematics tracers, including stars in the central regions, globular clusters in the intermediate regions, and satellite galaxies extending to the virial radius, indicate that the halo has a shallow density profile towards the center~\cite{Oldham:2015,Oldham:2016}. Ref.~\cite{DeLaurentis:2022nrv} fits the data in~\cite{Oldham:2015,Oldham:2016} with a cored Burkert profile~\cite{Burkert:1995yz} and shows that the halo has a $90~{\rm kpc}$ core, in contrast to a NFW cusp. Such a density core could be produced in SIDM models due to collisional thermalization~\cite{Rocha:2012jg,Kaplinghat:2015aga,Sagunski:2020spe}. In this work, we revisit the M87 constraints on dark matter annihilations based on the M87 halo model in~\cite{DeLaurentis:2022nrv}. We note that Ref.~\cite{Alvarez:2020fyo} adopts the same NFW halo for M87 as in~\cite{Lacroix:2015lxa} for the smooth part, while assuming an SIDM spike $\rho_{\rm sp}\propto r^{-7/4}$. With the cored Burkert halo, we expect the constraints will be even weaker than those in~\cite{Alvarez:2020fyo}. Moreover, to be concrete, we consider a dark matter model with a light mediator, where a fermionic dark matter particle interacts with scalar or vector bosons, which could decay to Standard Model fermions consequentially; see~\cite{Boehm:2003hm,Pospelov:2007mp,Arkani-Hamed:2008hhe,Feng:2010zp}. The model is well motivated for this work, as the light mediator could naturally induce strong dark matter self-interactions~\cite{Feng:2009hw, vandenAarssen:2012vpm,Tulin:2013teo} that produce the density core in the M87 halo. We will take into account the Sommerfeld enhancement effect~\cite{Hisano:2004ds, Cassel:2009wt, Slatyer:2009vg} in calculating the $J$ and $Q$ factors for s- and p-wave annihilations. With the Burkert halo model for M87, we will show that the smooth part of the halo completely dominates the contribution to annihilation signals and the spike's contribution is negligible. The projected gamma-ray fluxes are several orders of magnitude below the limits derived from~\cite{Lacroix:2015lxa}. This conclusion also holds if we take a spike profile of $\rho_{\rm sp}\propto r^{-9/4}$ if the density core is generated by astrophysical processes, while dark matter is collisionless.

The rest of the paper is organized as follows. In Sec.~\ref{sec:density}, we discuss the cored halo of M87 and the density spike surrounding its SMBH, as well as the velocity dispersion and escape velocity profiles of dark matter particles. In Sec.~\ref{sec:model}, we present the dark matter model with a light mediator. In Sec.~\ref{sec:jq}, we evaluate the $J$ and $Q$ factors for dark matter annihilations. In Sec.~\ref{sec:constraints}, we show the projected gamma-ray fluxes and compare them with observational constraints. In Sec.~\ref{sec:SIDM}, we derive an SIDM halo model for M87 that incorporates the effects of the baryonic potential and compute the corresponding fluxes. We conclude in Sec.~\ref{sec:con}.

\section{Dark Matter Spike of M87}
\label{sec:density}

We follow~\cite{DeLaurentis:2022nrv} and assume that the halo of M87 follows a Burkert profile~\cite{Burkert:1995yz},
\begin{equation}
    \rho_{\rm B}(r)=\frac{\rho_{0}}{\left(1+r/r_{0}\right)[1+({r}/{r_{0}})^2]},
\label{eq:burkert}
\end{equation}
where $\rho_{0}$ is the central density and $r_{0}$ is the scale radius. We take their best-fit values $\rho_{0}=6.94\times 10^6~{\rm M_{\odot}~kpc^{-3}}$ and $r_{0}=91.2~{\rm kpc}$ from~\cite{DeLaurentis:2022nrv}. The virial radius is $R_{200}=1.3~{\rm Mpc}$ and the total halo mass is $1.3\times10^{14}~{\rm M_\odot}$. In the limit of $r\rightarrow0$, the density approaches to the constant $\rho_0$, in contrast to an NFW cusp $\rho_{\rm NFW}\propto r^{-1}$. 

The mass of the M87 SMBH is $M_{\bullet}\approx6.5\times 10^9~{\rm M_{\odot}}$~\cite{EventHorizonTelescope:2019dse} and its radius of influence can be estimated as 
\begin{equation}
    r_{\rm in}= \frac{GM_{\bullet}}{\sigma^2_{\rm h}\left(r_{\rm in}\right)},
\label{eq:rin}
\end{equation}
where $G$ is the Newton constant and $\sigma_{\rm h}\left(r_{\rm in}\right)$ is the velocity dispersion of dark matter particles at $r=r_{\rm in}$. We determine the velocity dispersion profile $\sigma_{\rm h}(r)$ through the Jeans equation~\cite{Binney:2008},
\begin{equation}
   \sigma_{\rm h}^2\left(r\right)=\frac{1}{\rho_{\rm B}\left(r\right)}\int_r^{\infty}dr' \rho_{\rm B}\left(r'\right) \frac{d \Phi}{dr'}.
\label{eq:jeans}
\end{equation}  
The potential gradient is $d\Phi/dr=G\left[M_{\star}\left(r\right)+M_{h}\left(r\right)\right]/r^2$, where $M_{\star}\left(r\right)$ and $M_{h}\left(r\right)$ are stellar and halo masses profiles, respectively. Following~\cite{DeLaurentis:2022nrv}, we adopt the Nuker mass profile to model the stellar distribution of M87~\cite{Baes_2020}, where the surface luminosity within $210~{\rm kpc}$ is parameterized as 
\begin{equation}
   I\left(R\right)=I_0 \left(\frac{R}{R_{\star}}\right)^{-\zeta}\left[1+\left(\frac{R}{R_{\star}}\right)^{\alpha}\right]^{\frac{\zeta-\eta}{\alpha}},
\label{eq:nuker}
\end{equation}
where $I_0=3.5\times 10^9~{\rm L_{\odot}~ kpc^{-2}}$, $R_{\star}=1.05 ~{\rm kpc}$, $\zeta=0.186$, $\eta=1.88$, and $\alpha=1.27$. Assuming a constant light-to-mass ratio, $M_{\rm sph}/L_{\rm B}$, the deprojected 3D stellar density profile is 
\begin{equation}
   \rho_{\star}\left(r\right)=-\left(\frac{M_{\rm sph}}{L_\star}\right)\frac{1}{\pi}\int_r^{\infty}\frac{dI}{dR}\frac{dR}{\sqrt{R^2-r^2}}.
\label{}
\end{equation}
where we take $M_{\rm sph}/L_\star=8.6 ~{\rm M_{\odot} L_{\odot}^{-1}}$~\cite{DeLaurentis:2022nrv}. The stellar mass profile is calculated as $M_\star(r)=4\pi\int\rho_\star(r') {r'}^2dr'$. 

We solve Eq.~\ref{eq:rin} iteratively and find the radius of influence to be $r_{\rm in}\approx27.8~{\rm pc}$ after including both halo and stellar contributions when calculating the velocity dispersion profile in Eq.~\ref{eq:jeans}. If we included only the halo, but not the stars, the velocity dispersion would decrease by a factor of $2$, and the radius of influence would increase to $r_{\rm in}\approx136~{\rm pc}$ accordingly. 

The presence of the SMBH at the center of M87 induces a dark matter density spike and its profile scales as $r^{-\gamma_{\rm sp}}$ within $r_{\rm in}$. In the case of dark matter self-interactions, $\gamma_{\rm sp}=\left(3+a\right)/4$~\cite{Shapiro:2014oha}, where the parameter $a$ characterizes the velocity dependence of the self-scattering cross section of dark matter particles, $\sigma \propto v^{-a}$. Thus the density profile of the spike is 
\begin{equation}
    \rho_{\rm sp}=\rho_{\rm B}\left(r_{\rm sp}\right)\left(\frac{r}{r_{\rm sp}}\right)^{-\frac{3+a}{4}},
\label{eq:spike}
\end{equation}
where we have set the spike radius to be the radius of influence of the SMBH, $r_{\rm sp}=r_{\rm in}$. In our numerical analysis, we will fix $a=4$, which is the maximum value expected from dark matter self-interactions mediated by a light mediator through a Yukawa potential~\cite{Feng:2009hw,Tulin:2013teo}, and hence $\gamma_{\rm sp}=7/4$. For comparison, $\gamma_{\rm sp}=(9-2\gamma)/(4-\gamma)$ in the case of collisionless dark matter (CDM)~\cite{Gondolo:1999ef}, where $\gamma$ is the logarithmic density slope of the smooth part of the inner halo $\rho\propto r^{-\gamma}$. For an NFW halo, $\gamma=1$ and $\gamma_{\rm sp}=7/3$. For a cored CDM halo, $\gamma=0$ and $\gamma_{\rm sp}=9/4$. Thus the SIDM spike is less steep, compared to the CDM case. We note that the power-law slope for the CDM spike presented above assumes that the black hole grows adiabatically and is unaffected by environmental disruptions. However, studies have shown that processes such as mergers~\cite{Merritt:2002vj} and dynamical heating from interactions with stellar clusters~\cite{Gnedin:2003rj} can soften the spike slope. For example, in the latter case, the slope becomes $\gamma_{\rm sp}=3/2$ for CDM~\cite{Gnedin:2003rj}; see also~\cite{Balaji:2023hmy,Christy:2023tdv} for related discussions.

In the central region of the halo, dark matter particles can plunge into the black hole once they pass the marginally-bound circular orbit with the total energy $E=m_{\rm \chi}$ in the Schwarzschild geometry~\cite{Sadeghian:2013laa, Chandrasekhar:1985kt, Shapiro:1983du}. The marginally-bound radius is 
\begin{equation}
    r_{\rm m}=2 \,R_\bullet,
\label{}
\end{equation}
where $R_\bullet=2\,GM_{\bullet}$ is the Schwarzschild radius. For the M87 black hole, $M_{\bullet}=6.5\times 10^9~{\rm M_{\odot}}$, and hence $r_{\rm m}=1.24\times 10^{-3} ~{\rm pc}$. As a result, the dark matter density is zero for radii less than $r_{\rm m}$. Therefore, we assume that the full dark matter density profile follows a piecewise function as 
\begin{equation}
    \rho_{\rm}\left(r\right) =
	\left
	\{ \begin{array}{cc} 
	0 & r < r_{\rm m} \\
	\rho_{\rm sp}\left(r\right) & r_{\rm m}\leq r < r_{\rm sp}\\ 
        \rho_{\rm B}\left(r\right) & r_{\rm sp}\leq r \leq R_{200} 
	\end{array}.
	\right.
\label{eq:fulldensity}
\end{equation}  

 \begin{figure}[t!]
    \includegraphics[scale=0.203]{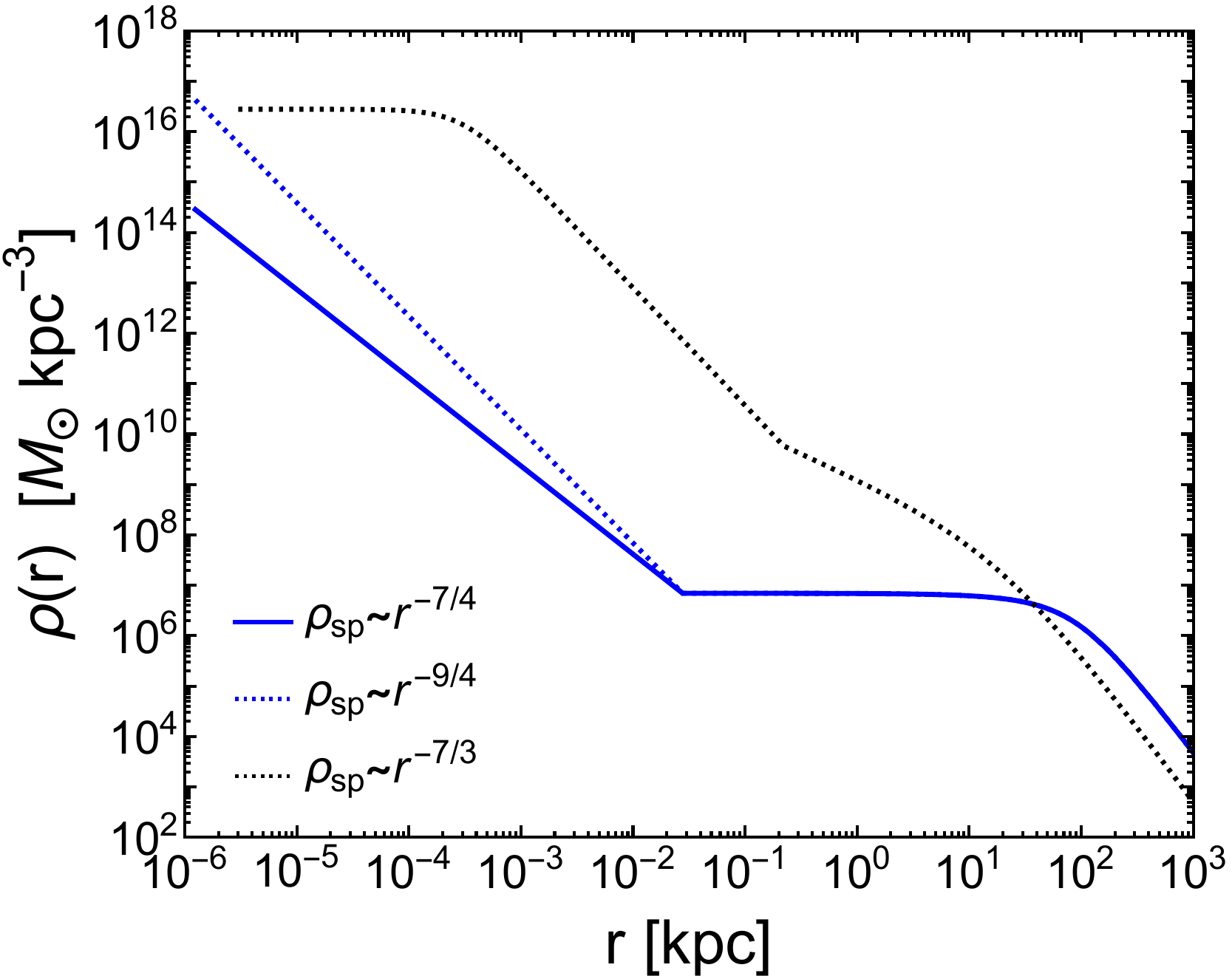}
    \includegraphics[scale=0.2]{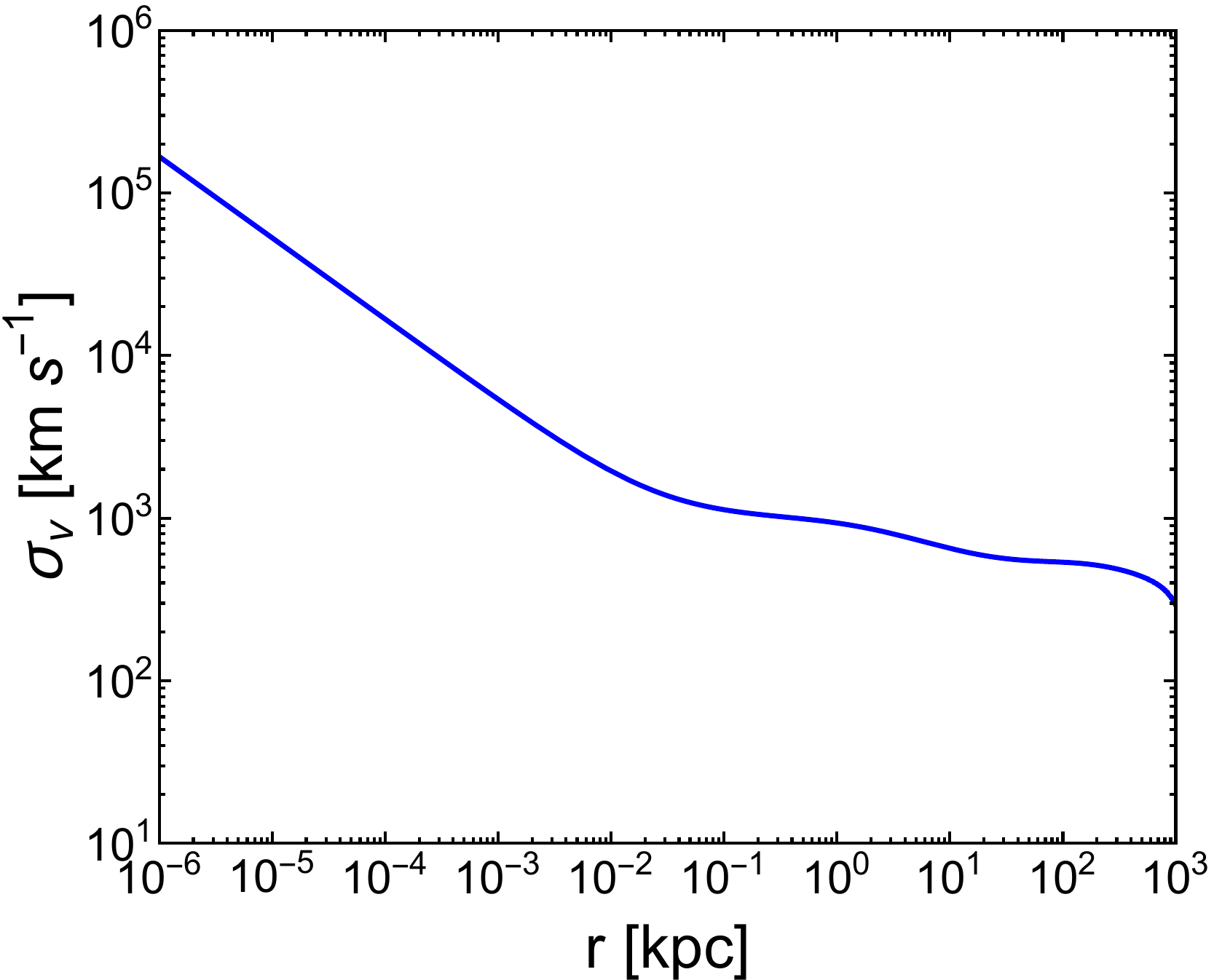} 
    \includegraphics[scale=0.2]{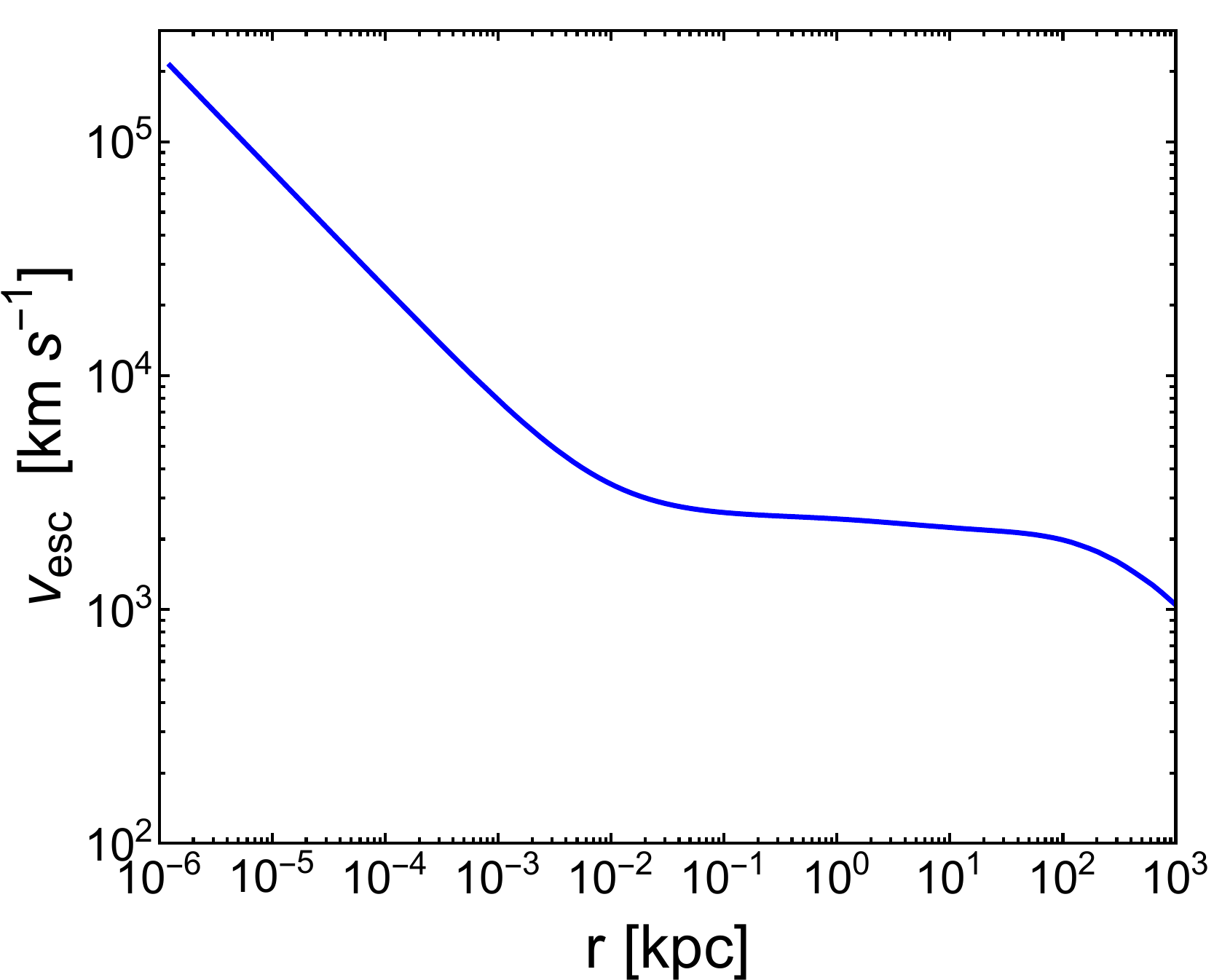}  
    \caption{Left: Dark matter density profile of M87 with the $\rho_{\rm sp}\propto r^{-7/4}$ motivated by dark matter self-interactions (blue-solid). For comparison, we also show a CDM profile adapted in Ref.~\cite{Lacroix:2015lxa} (dotted-black) and a profile with the spike $\rho_{\rm sp}\propto r^{-9/4}$ (dotted-blue), which assume a CDM halo, but having a cored profile for the smooth part. Middle: Velocity dispersion profile of dark matter particles in M87. Right: Escape velocity profile.} 
    \label{fig:profiles}
\end{figure}

In Fig.~\ref{fig:profiles} (left panel), we show the full dark matter density profile of M87 with the spike (solid-blue). In the spike, the density scales as $\rho(r)\propto r^{-7/4}$ as we have set $a=4$ assuming the self-scattering cross section scale as $\propto v^{-4}$. For comparison, we also show the density profile of M87 that Ref.~\cite{Lacroix:2015lxa} assumed to derive the constraints on dark matter annihilation (dotted-black). It includes an annihilation plateau towards the central region. We see that the spike density we consider can be a factor of $10^5$ smaller than that adapted in~\cite{Lacroix:2015lxa}, and hence we expect the constraints on dark matter annihilation are much weaker with our halo model. Ref.~\cite{Lacroix:2015lxa} assumed a NFW halo model for M87
\begin{equation}
    \rho_{\rm NFW}=\frac{\rho_{s}}{(r/r_s)(1+r/r_s)^2},
\end{equation}
with a scale density of $\rho_s=6.57\times 10^7~{\rm M_{\odot} kpc^{-3}}$ and a scale radius of $r_s=20~{\rm kpc}$. The spike density slope is $\gamma_{\rm sp}=7/3$, i.e., $\rho_{\rm sp}\propto r^{-7/3}$. The cored Burkert halo model proposed in~\cite{DeLaurentis:2022nrv} has much smaller  densities for the smooth part within $r\lesssim10~{\rm kpc}$. Additionally, we show a spike density profile $\rho_{\rm sp}\propto r^{-9/4}$ assuming a CDM halo (dotted-blue), but with a cored Burkert profile for the smooth part. 

For the dark matter model we will consider, there is a light mediator and it induces the Sommerfeld enhancement effect, which depends on the relative velocity of dark matter particles. In order to estimate the annihilation rate, we first evaluate the velocity dispersion and escape velocity of dark matter particles in the M87 halo. In Eq.~\ref{eq:jeans}, we derived the velocity dispersion $\sigma_{\rm h}(r)$ at the radius $r$ using the Jeans equation with the contributions from both dark and luminous matter for radii $r>r_{\rm sp}$. Now we further include the SMBH and the spike in calculating $d\Phi/dr$ in Eq.~\ref{eq:jeans} to obtain the velocity dispersion profile $\sigma_{\rm v}(r)$ for the whole range, as shown in Fig.~\ref{fig:profiles} (middle panel). Note within the spike radius we have $\sigma_{\rm v}(r)\propto r^{-1/2}$ as expected~\cite{Shapiro:2014oha}. The velocity dispersion increases from $500~{\rm km~s^{-1}}$ to $1000~{\rm km~s^{-1}}$ as the radius decreases from the outer halo $100~{\rm kpc}$ to the spike radius $30~{\rm pc}$. If we only include the halo component, the velocity dispersion is a constant within the region $\sigma_{\rm v}\approx500~{\rm km~{s^{-1}}}$. Thus the presence of the stellar component can increase the dispersion by a factor of $2$ before reaching the spike.

For the escape velocity, we calculate it in the Newtonian limit $v_{\rm esc}\left(r\right)=\sqrt{2 \left[\Phi\left(\infty\right)-\Phi\left(r\right)\right]}$, where the gravitational potential is given by
\begin{equation}
\Phi\left(r \right)= -4\pi G\left[\frac{1}{r}\int_0^r dr' r'^2 (\rho_{\rm}(r')+\rho_{\rm \star}(r'))+\int_r^\infty dr' r' (\rho_{\rm}\left(r')+\rho_{\rm \star}(r')\right)\right]-\frac{G M_{\bullet}}{r},
\label{eq:potential} 
\end{equation}
and we will set $\Phi\left(\infty\right)=0$~\cite{Binney:2008}. In Fig.~\ref{fig:profiles} (right panel), we show the escape velocity profile. The escape velocity increases very mildly from $2000~{\rm km~{s^{-1}}}$ to $2100~{\rm km~{s^{-1}}}$ as the radius decreases from $100~{\rm kpc}$ to $0.03~{\rm kpc}$. For $r\lesssim r_{\rm sp}\approx30~{\rm pc}$, the escape velocity increases as $v_{\rm esc}\propto r^{-1/2}$ due to the presence of the SMBH.

\section{Dark matter model with a light mediator}
\label{sec:model}
As an example, we consider the dark matter model with a light mediator and the interaction term is described by the following Lagrangian 
\begin{equation}
    \mathcal{L}_{\rm int} =
	\left
	\{ \begin{array}{cc} 
	g_{ \chi}  \bar{\chi} \gamma^{\rm \mu}\chi \phi_{\rm \mu} & {\rm vector ~ mediator}\\
	&\\
	g_{ \chi}  \bar{\chi} \chi \phi & {\rm scalar ~mediator}
	\end{array},
	\right.
\label{eq:lagrangian}
\end{equation}
where dark matter $\chi$ is a fermion, $\phi_\mu$ and $\phi$ represent vector and scalar mediators, respectively, and $g_\chi$ is the interaction strength. For the vector mediator, we assume that it couples to the fermions of the standard model via kinetic mixing with the photon~\cite{Holdom:1985ag}. For the scalar mediator, $\phi$ couples to the fermions via Higgs mixing~\cite{Patt:2006fw}. In both cases, a pair of dark matter particles can annihilate to two mediator particles, which subsequently decay into two standard model fermions. In our numerical study, we will focus on the $e\bar{e}e\bar{e}$ and $\mu\bar{\mu}\mu\bar{\mu}$ final states~\cite{Pospelov:2007mp,Arkani-Hamed:2008hhe,Pospelov:2008jd}. Note that the dark matter model characterized by the Lagrangian in Eq.~\ref{eq:lagrangian} has been extensively studied in the literature, including in the contexts of direct and indirect detection as well as cosmological constraints; see, e.g.,~\cite{Kaplinghat:2013yxa,Bernal:2015ova,Bringmann:2016din,Cirelli:2016rnw,Chu:2016pew,Baldes:2017gzu,Kahlhoefer:2017umn,Duerr:2018mbd,Hufnagel:2018bjp,PandaX-II:2018xpz,Bernal:2019uqr,Wu:2022wzw,Alvarez:2023fjj,PandaX:2023xgl,Xu:2024iny}. In this work, we do not aim to comprehensively constrain the model; rather, we use it as an example to demonstrate how the constraints from M87 can be relaxed when adopting a cored halo model. We refer interested readers to the relevant references for further discussion on other aspects of the model.

The tree-level annihilation cross sections for $\chi \bar{\chi} \xrightarrow[]{}\phi \phi$ with vector and scalar mediators are
\begin{equation}
    (\sigma v)^{\rm tree} =
	\left
	\{ \begin{array}{cc} 
	\frac{\pi \alpha_{\rm \chi}^2}{m_{\rm \chi}^2}\,\sqrt{1-\frac{m_{\rm \phi}^2}{m_{\rm \chi}^2}} & {\rm vector ~mediator~(s-wave)}\\
	&\\
	\frac{3}{4}\frac{\pi \alpha_{\rm \chi}^2}{m_{\rm \chi}^2} \, v^2\, \sqrt{1-\frac{m_{\rm \phi}^2}{m_{\rm \chi}^2}} & {\rm scalar ~ mediator~(p-wave)}
	\end{array},
	\right.
\label{eq:AnnihilationCrossSection}
\end{equation}
respectively, where $\alpha_{\rm \chi}=g_{\rm \chi}^2/4\pi $, $m_{\rm \chi }$ is the particle mass of dark matter, $m_{\rm \phi }$ is the mediator mass, and $v=\left|\Vec{v}_1-\Vec{v}_2\right|$ is the relative velocity between two annihilating dark matter particles with velocities $\Vec{v}_1$ and $\Vec{v}_2$; see, e.g.,~\cite{Tulin:2013teo}. For the model we consider, the corresponding s-wave and p-wave Sommerfeld enhancement factors are~\cite{Cassel:2009wt}
\begin{equation}
    S_{n}\left(v\right) =
	\left
	\{ \begin{array}{cc} 
	\frac{\pi}{a}\frac{\sinh{\left(2\pi a b\right)}}{\cosh{\left(2\pi a b\right)}-\cos{\left(2\pi\sqrt{b-(a b)^2}\right)}} & {n=0 ~({\rm s-wave})}\\
	&\\
	\frac{\pi}{a}\frac{\sinh{\left(2\pi a b\right)}}{\cosh{\left(2\pi a b\right)}-\cos{\left(2\pi\sqrt{b-(a b)^2}\right)}}\frac{(b-1)^2+4(a b)^2}{1+4(a b)^2} & { n=2 ~{\rm (p-wave)}}
	\end{array},
	\right.
\label{eq:sommerfeld}
\end{equation}
respectively, where $a=v/2\alpha_{\rm \chi}$, $b=6\alpha_{\rm \chi}m_{\rm \chi}/\pi^2m_{\rm \phi}$, and $n$ indicates the power which the annihilation cross section depends on the relative velocity. We collectively write the s- and p-wave annihilation cross sections as $(\sigma v)=v^n S_{ n}\left(v\right)\times (\sigma v)_0$, where $\left(\sigma v\right)_0$ represents the tree-level annihilation cross section excluding the $v^2$ term (for the p-wave case).

\section{Velocity-Averaged J and Q factors}
\label{sec:jq}

The differential photon flux due to the pair annihilation of dark matter particles is 
\begin{equation}
    \frac{d^2\Psi}{dE_{\rm \gamma} d\Omega}\;=\;\frac{1}{2}\,\frac{\left(\sigma v\right)_0}{4\pi m_{\rm \chi}^2}\,\frac{dN_{\rm \gamma}}{dE_{\rm \gamma}}\left(J\left(\Omega\right)+Q\left(\Omega\right)\right) ,
\label{eq:flux}    
\end{equation}
where $dN_{\rm \gamma}/dE_{\rm \gamma}$ is the energy spectrum of the photon produced per annihilation, see, e.g.,~\cite{Cirelli:2010xx}, and $(\sigma v)_0$ is the velocity-independent part of the annihilation cross section, $J(\Omega)$ and $Q(\Omega)$ characterize contributions from the smooth halo and density spike, respectively. For the $J(\Omega)$ and $Q(\Omega)$ factors, we evaluate them as 
\begin{equation}
    J\left(\Omega\right)=\int d\ell\int d^3v_1 \int d^3 v_2 \,  v^n\,S_{n}(v)f\left(r,\Vec{v}_1\right) f\left(r,\Vec{v}_2\right)\rho^2_{\rm B}(r),
\label{eq:jfactor}    
\end{equation}
\begin{equation}
    Q\left(\Omega\right)=\frac{1}{D^2}\int_{r_{\rm m}}^{r_{\rm sp}} dr\, r^2 \int d^3v_1 \int d^3 v_2   \, v^n\,S_{n}(v) f\left(r,\Vec{v}_1\right) f\left(r,\Vec{v}_2\right) \rho_{\rm sp}^2(r),
\label{eq:qfactor}    
\end{equation}
where we have followed~\cite{Boddy:2019qak} and included the velocity-dependent part of the annihilation cross section into the evaluation of the $J$ and $Q$ factors. $f(r,v)$ is the velocity distribution function of dark matter particles; $\ell$ is the line-of-sight distance, which is related to the distance from the center of M87 ($r$) and the distance from Earth to the halo center of M87 ($D$) as $r^2 = \ell^2 + D^2 - 2 \ell D \cos{\theta}$. We assume that the distribution function $f(r, \vec{v_1})$ follows an isotropic Maxwell-Boltzmann distribution with a truncation at the escape velocity $f(r, \vec{v_1})\propto\exp[-v^2_1(r)/2\sigma^2_{\rm v}(r)]$, where $\sigma_{\rm v}(r)$ is the 1D velocity dispersion at radius $r$. For a given function of $g(v)$, we have
\begin{equation}
 N \int d^3v_1 \, d^3v_2 \, {e}^{-\frac{v_1^2}{2 \sigma^2_{\rm v}(r)}}  {e}^{-\frac{v_2^2}{2 \sigma^2_{\rm v}(r)}} g(v)=N\int_0^{2 v_{\rm esc}(r)} 4\pi v^2  {e}^{-\frac{v^2}{4 \sigma^2_{\rm v}(r)}}g(v) dv,
\label{}
\end{equation}
and $N$ is the normalization factor,
\begin{equation}
    N(r)=\frac{1}{\left[4 \pi \sigma^2_{\rm v}(r)\right]^{\frac{3}{2}}\left[{\rm erf}\left(\frac{v_{\rm esc}(r)}{\sigma_{\rm v}(r)}\right)-\frac{2}{\sqrt{\pi}}\frac{v_{\rm esc}(r)}{\sigma_{\rm v}(r)}{e}^{-\frac{v^2_{\rm esc}(r)}{\sigma^2_{\rm v}(r)}}\right]},
\label{}
\end{equation}
 where $v_{\text{esc}}(r)$ is the escape velocity at radius $r$ and ${\rm erf}(x)$ is the error function.

Putting these together, we can evaluate the $J$ and $Q$ factors as
\begin{align}
    J\left(\Omega\right)&= \int d\ell \, \rho^2_{\rm B}(r) N\left(r\right) \int_0^{2v_{\rm esc}} 4\pi v^{n+2}S_{ n}(v) {e}^{-\frac{v^2}{4\sigma_{\rm h}(r)^2}}dv,\label{eq:j}\\
    Q\left(\Omega\right)&=\frac{1}{D^2}\int_{r_{\rm m}}^{r_{\rm sp}} dr\, r^2 \rho_{\rm sp}(r)^2N(r)\int_0^{2v_{\rm esc}} 4 \pi v^{n+2}S_{n}(v) e^{-\frac{v^2}{4\sigma^2_{\rm v}(r)}} dv,
\label{eq:q}    
\end{align}
respectively, where $\rho_{\rm B}(r)$ and $\rho_{\rm sp}(r)$ are given in Eq.~\ref{eq:fulldensity}. We will take $\rho_{\rm sp}\propto\gamma^{-(3+a)/4}$ and $a=4$, the steepest logarithmic slope corresponding to Coulomb-like dark matter self-interactions.

\begin{figure}[t!]
	\includegraphics[scale=0.2]{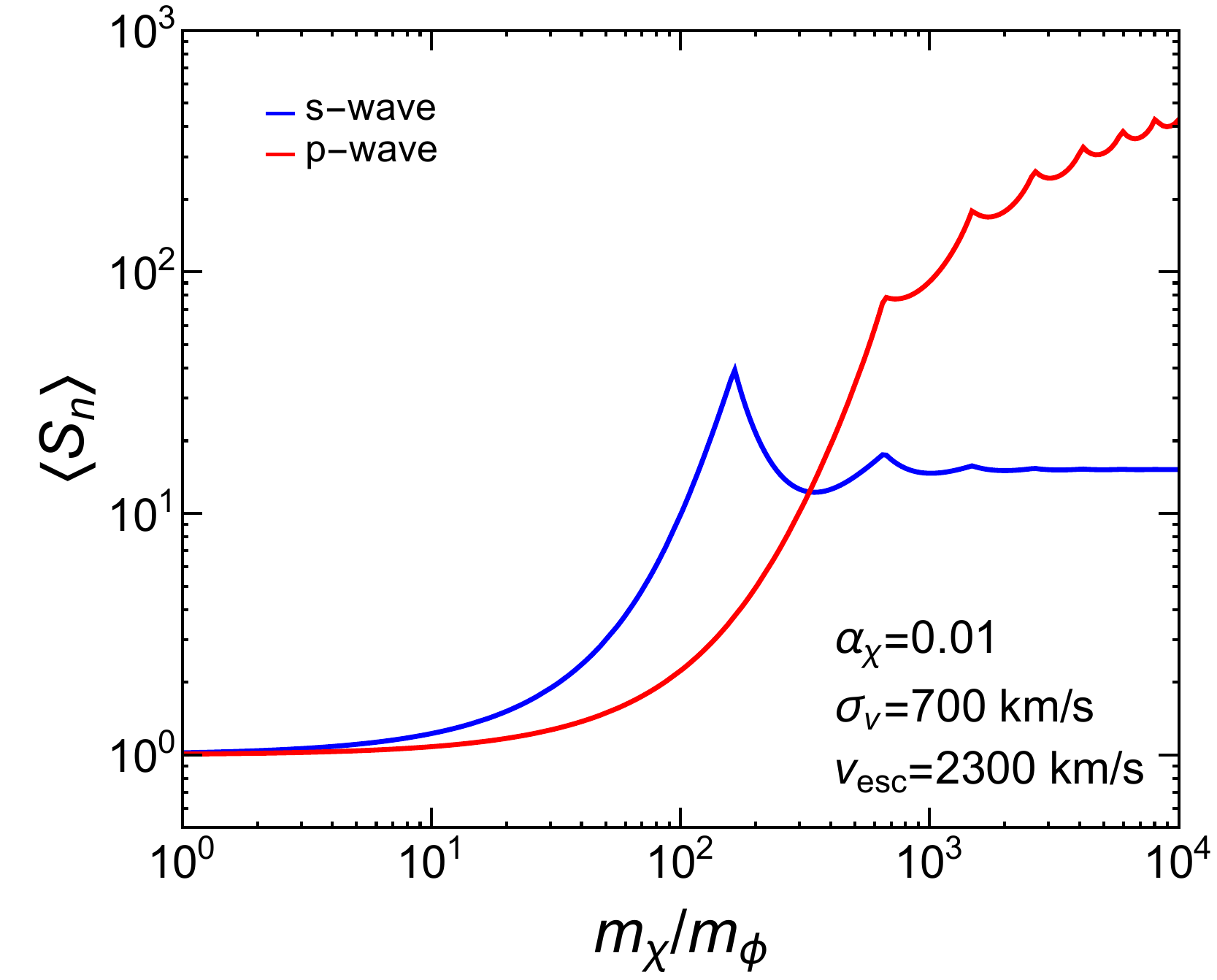}\;\;\;
	\includegraphics[scale=0.2]{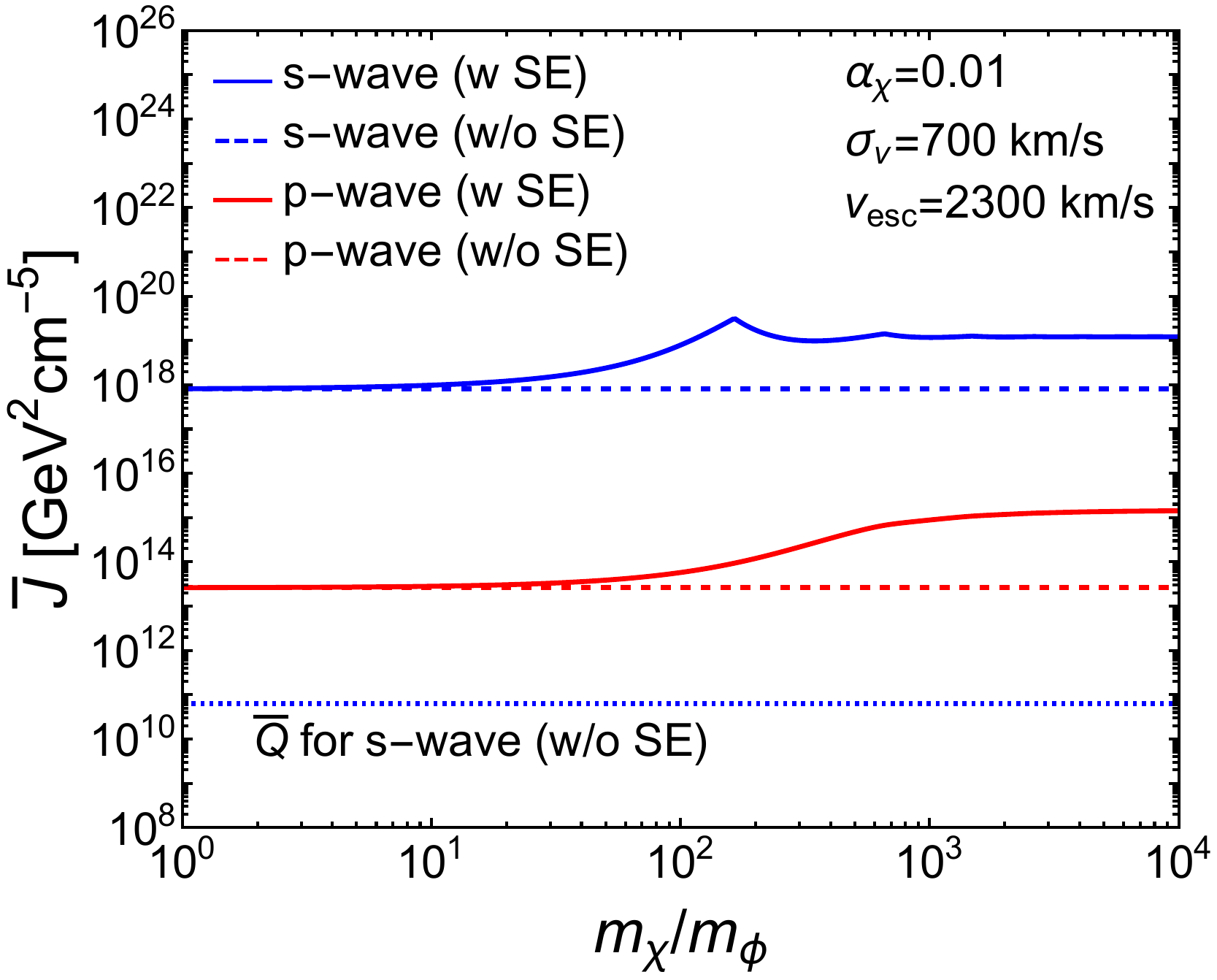}
	\includegraphics[scale=0.205]{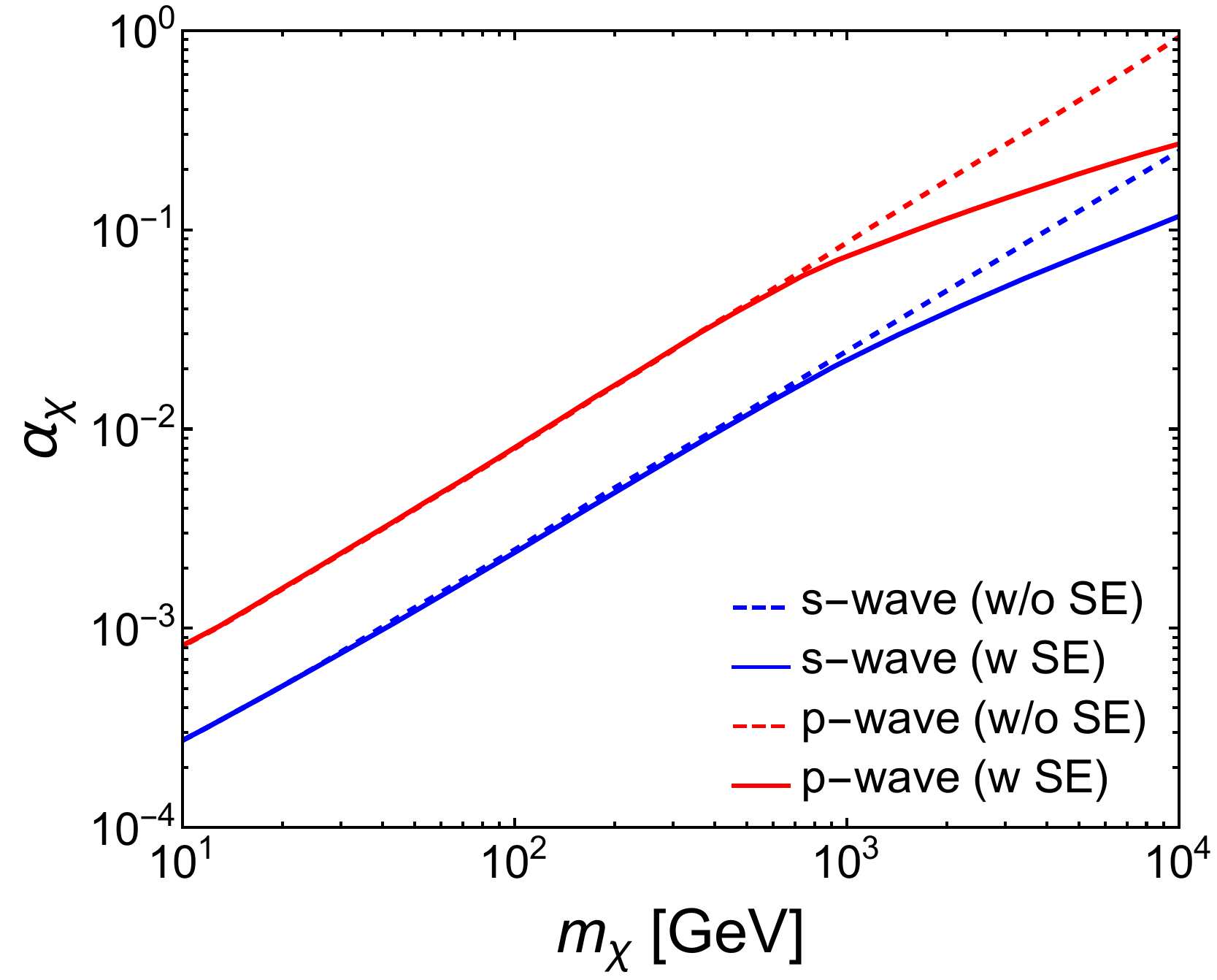}
	\caption{Left: Thermally-averaged s-wave (blue) and p-wave (red) Sommerfeld enhancement factors as a function of $m_{\rm \chi}/m_{\rm \phi}$, where $\alpha_{\rm \chi}=0.01$, $v_{\rm esc}=2300\, {\rm km\, s^{-1}}$, and $\sigma_{\rm v}=700\, {\rm km\, s^{-1}}$. Middle: Corresponding $\bar{J}$ values, with (solid) and without (dashed) taking into account the Sommerfeld enhancement effect. For comparison, we also show the s-wave $\bar{Q}$ value, without including enhancement effect (dotted). Right: Dark sector fine structure constant as a function of dark matter mass after imposing the relic density constraint for s-wave (blue) and p-wave (red) annihilations, with (solid) and without (dashed) taking into account the Sommerfeld enhancement effect in the relic abundance calculation. The mediator mass is fixed to $m_\phi=1~{\rm GeV}$ as an example.}
	\label{fig:factors}
\end{figure}

Fig.~\ref{fig:factors} (left panel) shows the thermally-averaged s-wave (blue) and p-wave (red) Sommerfeld enhancement factors as a function of $m_{\rm \chi}/m_{\rm \phi}$; see Eq.~\ref{eq:sommerfeld}. For illustration purposes, we have fixed $\alpha_{\rm \chi}=0.01$, $\sigma_{\rm v}=700~{\rm km~{s^{-1}}}$, and $v_{\rm esc}=2300~{\rm km~{s^{-1}}}$. As $m_\chi/m_\phi$ increases, the enhancement factor increases with oscillatory features as $m_\chi/\phi\gtrsim100$. For s-wave, $\left<S_0\right>$ saturates to $15$, while for p-wave, $\left<S_2\right>$ increases to $400$. Fig.~\ref{fig:factors} (middle panel) illustrates the impact of Sommerfeld enhancement on the $J$ factor integrated over the solid angle $\bar{J}=\int d\Omega J(\Omega)$ for s-wave (blue) and p-wave (red) annihilations. To perform the integration over the line-of-sight distance, we have taken $D=16.5~{\rm Mpc}$ from Earth to M87 and set the $\ell$ range to be $D-R_{200}\leq\ell\leq D+R_{200}$, where the virial radius is $R_{200}=1.3~{\rm Mpc}$, and $0\leq\theta\leq4.5^{\circ}$, consistent with the angular size of the M87 halo. For $m_\chi/m_\phi\gtrsim100$, the Sommerfeld effect can boost the $\bar{J}$ values by $10$ and $100$ for s-wave and p-wave annihilations, respectively. Overall, the s-wave $\bar{J}$ value is approximately $10^4$ times larger than the p-wave $\bar{J}$ value, as the latter is suppressed by the $v^2$ term.

We also find that the $Q$ factor is much smaller than the $J$ factor for the halo model adopted for M87. To demonstrate this, we consider s-wave annihilation and set the Sommerfeld factor to $1$ in Eqs.~\ref{eq:j} and~\ref{eq:q}. In this case, the $J$ and $Q$ factors depend solely on the dark matter density profile. We find $\bar{J} \sim 8\times 10^{17}~{\rm GeV^2~cm^{-5}}$. For the $Q$ factor, taking $r_{\rm sp}=27.8~{\rm pc}$ and $\rho_{\rm sp}\propto r^{-7/4}$, we have $\bar{Q} \sim 6\times 10^{10}~{\rm GeV^2 ~cm^{-5}}$, which is a factor of $10^7$ smaller than the $\bar{J}$ factor; see~Fig.~\ref{fig:factors} (middle panel, dotted-blue). Even if we neglect the stellar potential in calculating the velocity dispersion of dark matter particles and take the spike radius to be $r_{\rm sp}=136~{\rm pc}$, the resulting $\bar{Q}$ value is $\sim2\times 10^{13}~{\rm GeV^2~cm^{-5}}$, still negligible compared to $\bar{J}$. Thus, the smooth part of the halo dominates the contribution for dark matter annihilation signals, and the density spike near the M87 black hole is negligible. We expect that the indirect detection constraints from the M87 can be significantly weakened for the scenario we consider.

\section{Constraints from M87}
\label{sec:constraints}

This section presents the anticipated gamma-ray flux resulting from dark matter annihilation within M87 under the framework of the light mediator model as discussed in Sec.~\ref{sec:model}.  We take the particle mass of dark matter to be in the range $10~{\rm GeV} \leq m_{\rm \chi} \leq 10~{\rm TeV}$ and consider two values for the mediator mass $m_{\rm \phi}=2~{\rm MeV}$ and $1~{\rm GeV}$, corresponding to $\bar{e}e\bar{e}e$ and $\bar{\mu}\mu\bar{\mu}\mu$ final states, respectively. We consider both s- and p-wave annihilations and their tree-level cross sections are given in Eq.~\ref{eq:AnnihilationCrossSection} and Sommerfeld factors in Eq.~\ref{eq:sommerfeld}. For given $m_\chi$ and $m_\phi$ values, $\alpha_\chi$ is fixed by the relict density constraint. We follow the procedure discussed in~\cite{Feng:2010zp, Tulin:2013teo} to calculate the thermal relic abundance via the annihilation process $\chi\bar{\chi}\rightarrow\phi\phi$. Fig.~\ref{fig:factors} (right), we show $\alpha_\chi$ as a function of $m_\chi$ for s-wave (solid-blue) and p-wave (solid-red) annihilations. For comparison, we also show the corresponding $\alpha_\chi$ values without taking into account the Sommerfeld effect in the relic density calculation (dashed). For $m_\chi\gtrsim1~{\rm TeV}$, the Sommerfeld effect becomes significant and $\alpha_{\rm \chi}$ needs to be reduced in order to obtain the correct relic abundance.

The line-of-sight integration in the $J$ factor Eq.~\ref{eq:j} is performed over the range $D-R_{200}<\ell<D+R_{200}$, where $D=16.5~{\rm Mpc}$ and $R_{200}=1.3~{\rm Mpc}$. According to the Fermi-LAT data~\cite{Fermi-LAT:2009nqy}, the angular resolution depends on the gamma-ray energy, expressed as $\theta = 0.8^\circ \left(E_{\gamma}/{\rm GeV}\right)^{-0.8}$, with $E_{\gamma}$ representing the energy of photons within the range $0.2~{\rm GeV}\leq E_{\gamma} \leq 31.5~{\rm GeV}$. For the $\bar{e}e\bar{e}e$ and $\bar{\mu}\mu\bar{\mu}\mu$ final states, we calculate the photon energy spectrum $dN_{\rm \gamma}/dE_{\rm \gamma}$ using the public code developed in~\cite{Cirelli:2010xx}.

 \begin{figure}[t!]
    \includegraphics[scale=0.24]{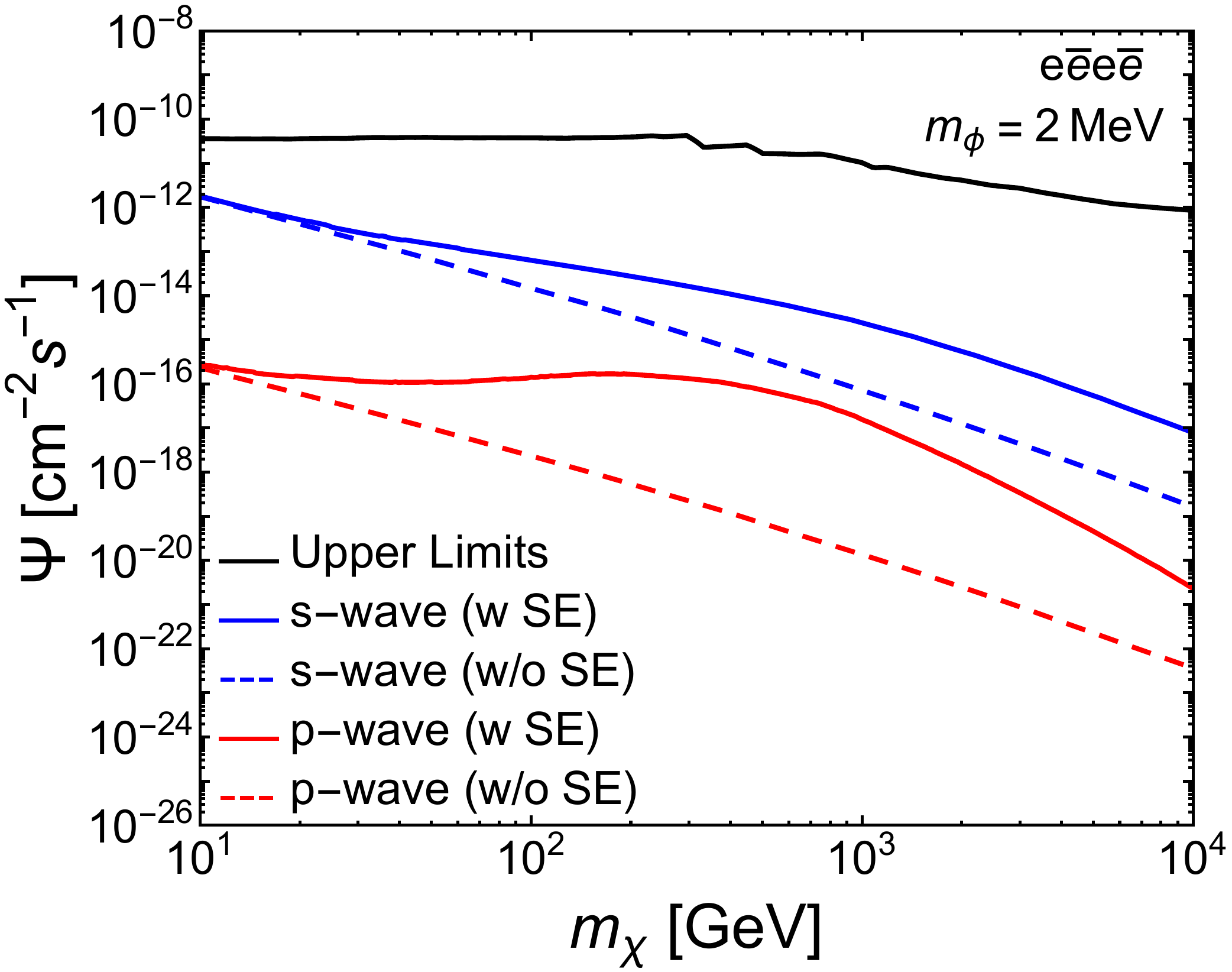}\;\;\;
    \includegraphics[scale=0.24]{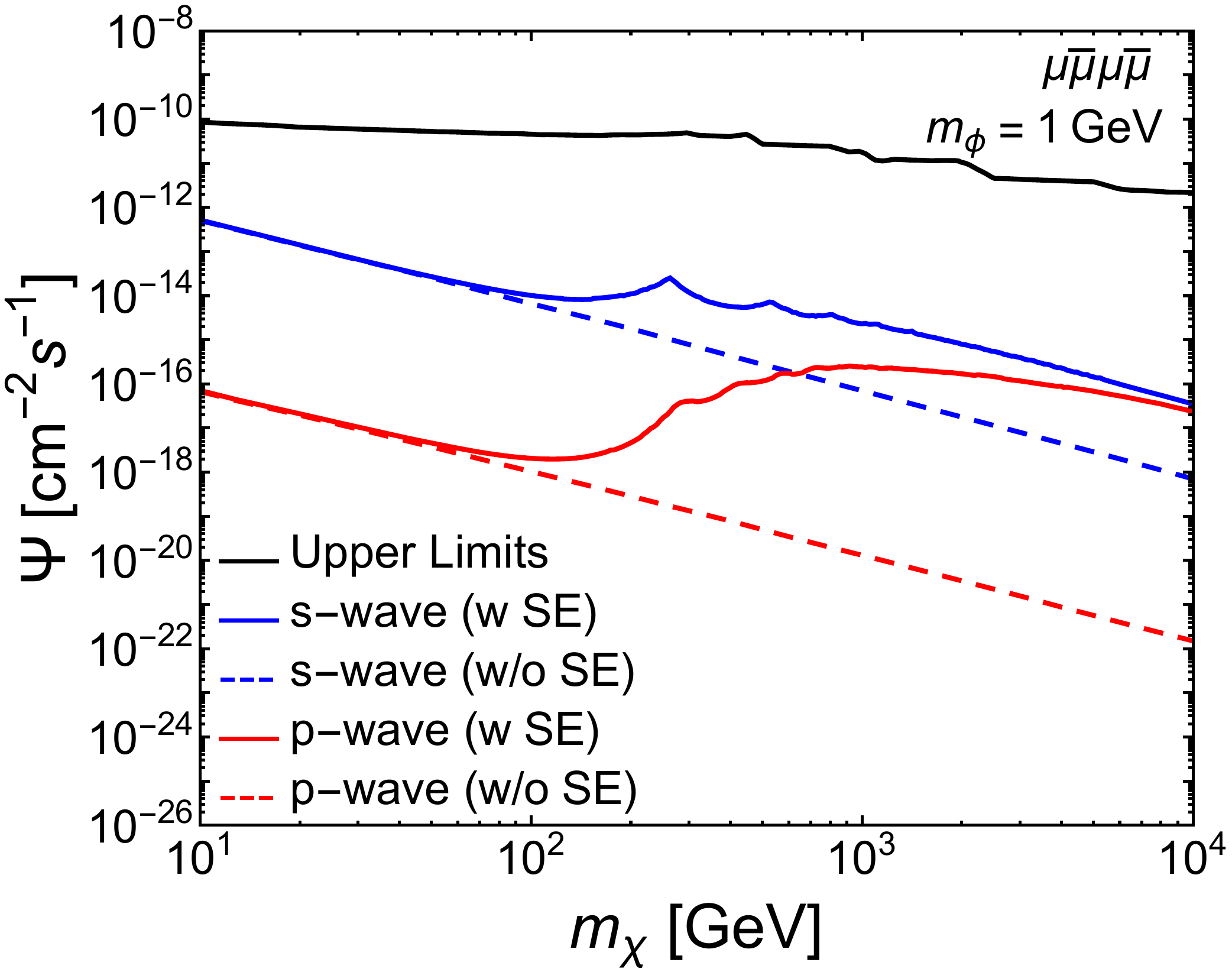}
    \caption{Left: Projected gamma-ray fluxes from s-wave (blue) and p-wave (red) dark matter annihilations $\chi\bar{\chi}\rightarrow\phi\phi\rightarrow e\bar{e}e\bar{e}$, with (solid) and without (dashed) taking into account the Sommerfeld enhancement effect. The mediator mass is fixed to $m_{\rm \phi}=2~{\rm MeV}$. Right: Similar to the left panel, but for $\chi\bar{\chi}\rightarrow\phi\phi\rightarrow \mu\bar{\mu}\mu\bar{\mu}$, where $m_{\rm \phi}=1~{\rm GeV}$. In both panels, the upper limits on the gamma-ray flux, which are derived from~\cite{Lacroix:2015lxa}, are shown for comparison (black).} 
    \label{F:Flux74}
\end{figure}

Fig.~\ref{F:Flux74} shows the predicted photon flux from the $e\bar{e}e\bar{e}$ (left panel) and $\mu\bar{\mu}\mu\bar{\mu}$ (right panel) final states for s-wave (blue) and p-wave (red) annihilations, with (solid) and without (dashed) taking into account the Sommerfeld enhancement effect, assuming the spike density as $\rho_{\rm sp}\propto r^{-7/4}$, motivated by dark matter self-interactions. For comparison, we show the flux limits (black), which are derived from the upper bounds on WIMP annihilation cross sections for $\bar{e}e$ and $\bar{\mu}{\mu}$ final states in~\cite{Lacroix:2015lxa}. Although the Sommerfeld effect boosts the signal strength for $m_\chi\gtrsim100~{\rm GeV}$, the fluxes predicted from our model are significantly below the upper limits, which are based on the Fermi-LAT measurements~\cite{Fermi-LAT:2009nqy}. 

As discussed in the previous section, for the cored halo model of M87, the smooth part of the halo dominates contributions to the annihilation signal, while the spike's contribution is negligible. For example, considering s-wave annihilation to the $\mu\bar{\mu}\mu\bar{\mu}$ final states with $m_{\rm \chi}=10~{\rm GeV}$ and $m_{\rm \phi}=1~{\rm GeV}$, the expected fluxes from the smooth part and the spike are $5\times 10^{-13}~{\rm cm^{-2}~s^{-1} }$ and $4\times 10^{-20}~{\rm cm^{-2}~s^{-1} }$, respectively. In the case of p-wave annihilation with $m_{\rm \chi}=10~{\rm GeV}$ and $m_{\rm \phi}=1~{\rm GeV}$, the fluxes from the smooth halo and spike are $7\times 10^{-17}~{\rm cm^{-2}~s^{-1} }$ and $9\times 10^{-20}~{\rm cm^{-2}~s^{-1} }$, respectively. We have further checked that for a density spike with $\rho_{\rm sp}\propto r^{-9/4}$ and $r_{\rm sp}=27.8~{\rm pc}$, motivated by a CDM halo, but with a cored smooth part of the halo, see Fig.~\ref{fig:profiles} (left, dotted-blue), the predicted fluxes are almost identical to those shown in Fig.~\ref{F:Flux74} assuming the $\rho_{\rm sp}\propto r^{-7/4}$ spike. This is not surprising as with the halo model we consider the spike has negligible contributions to annihilation signals, compared to the smooth halo. In contrast, taking the halo model in~\cite{Lacroix:2015lxa} for M87, we find that $\bar{Q}$ is $10^5$ times larger than $\bar{J}$, resulting in much stronger bounds on dark matter annihilations. Additionally, we have calculated the expected fluxes from the entire halo by integrating over $\theta$ from $0^\circ$ to $4.5^\circ$, representing a scenario without the limitation of the Fermi-LAT resolution, and found that they could increase by up to a factor of 4 toward $m_\chi = 10~{\rm TeV}$, while still remaining many orders of magnitude below current limits.

\section{Self-interacting dark matter halo model}
\label{sec:SIDM}

\begin{figure}[t]
    \centering
    \includegraphics[width=0.33\linewidth]{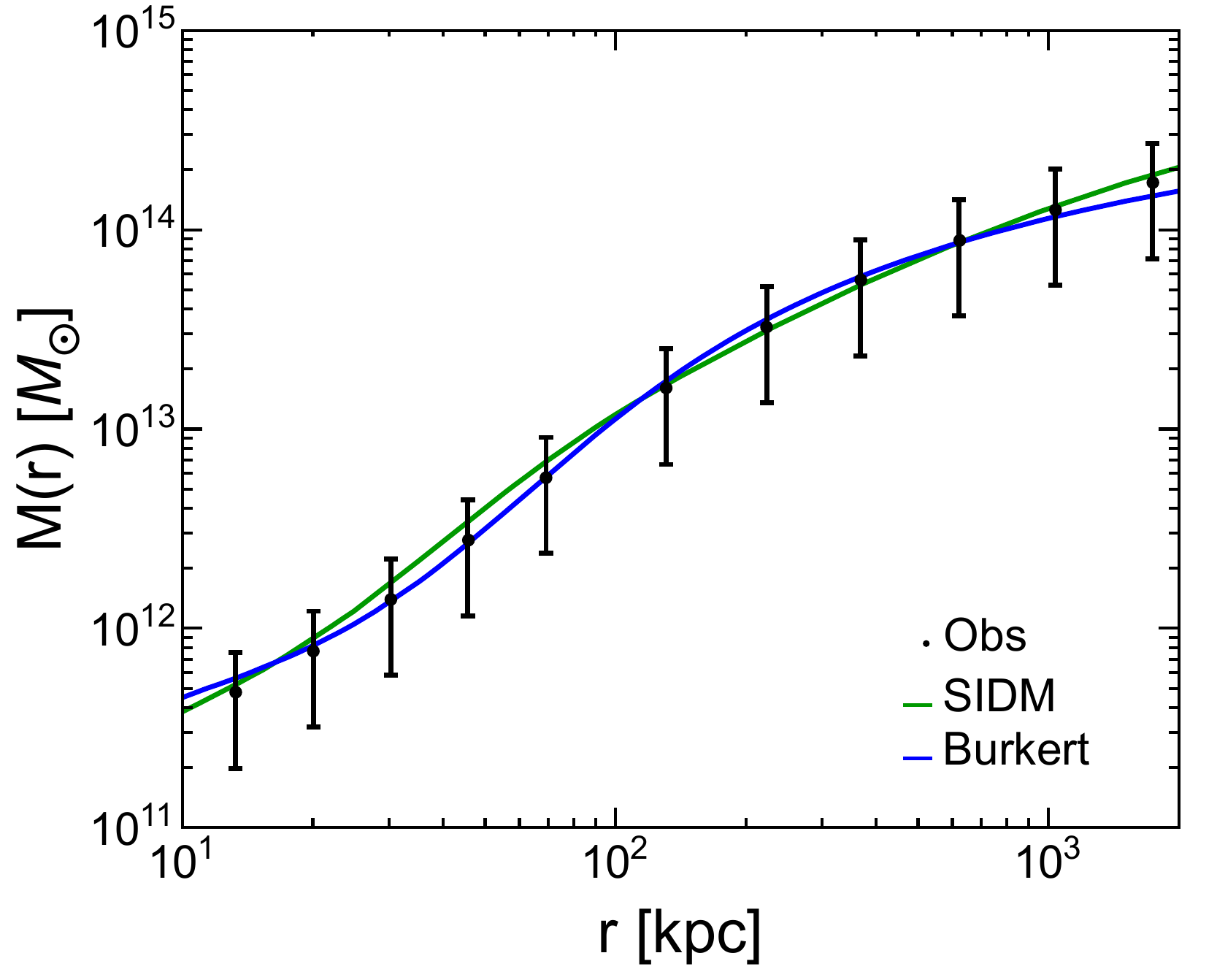}
     \includegraphics[width=0.33\linewidth]{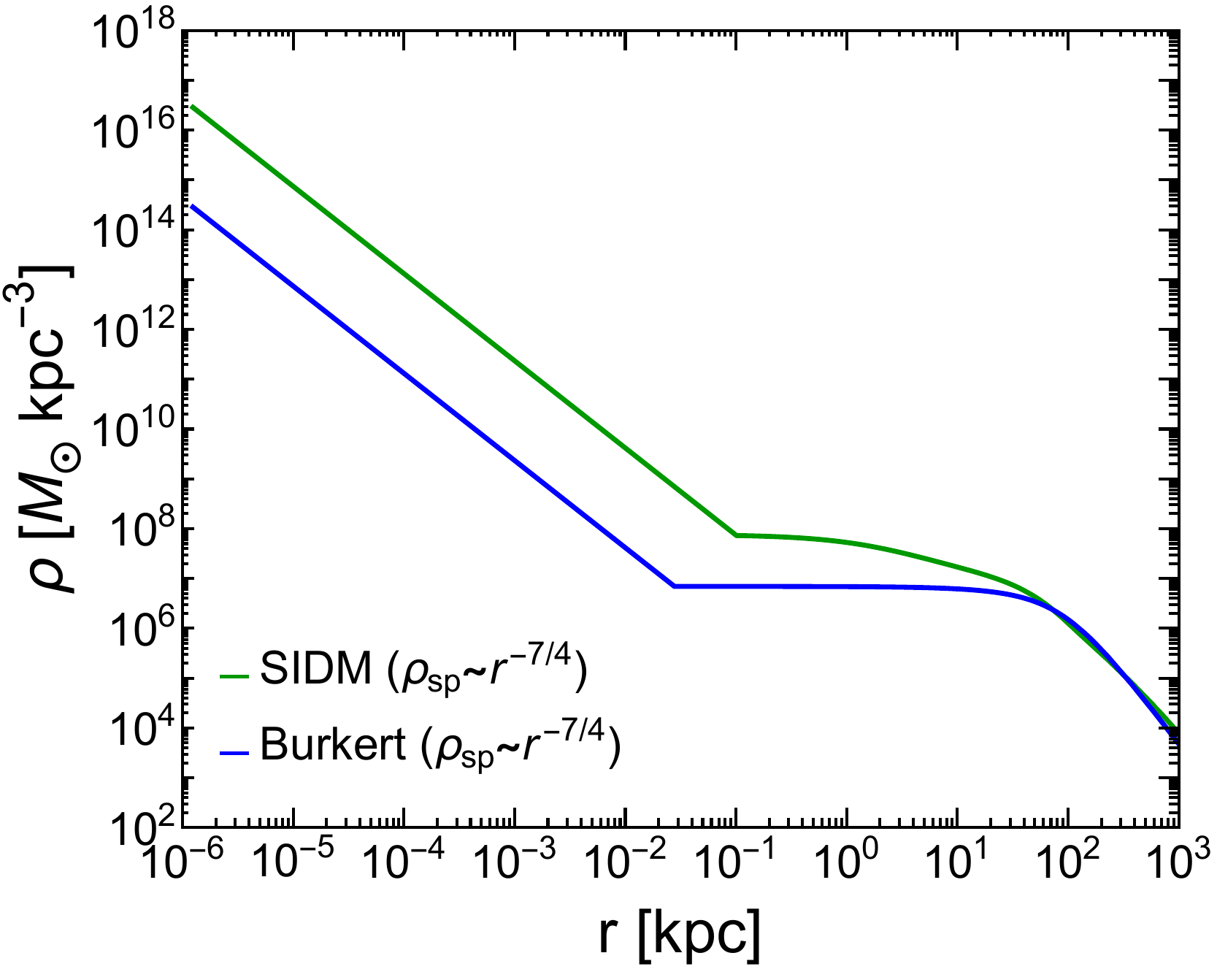}
        \includegraphics[width=0.33\linewidth]{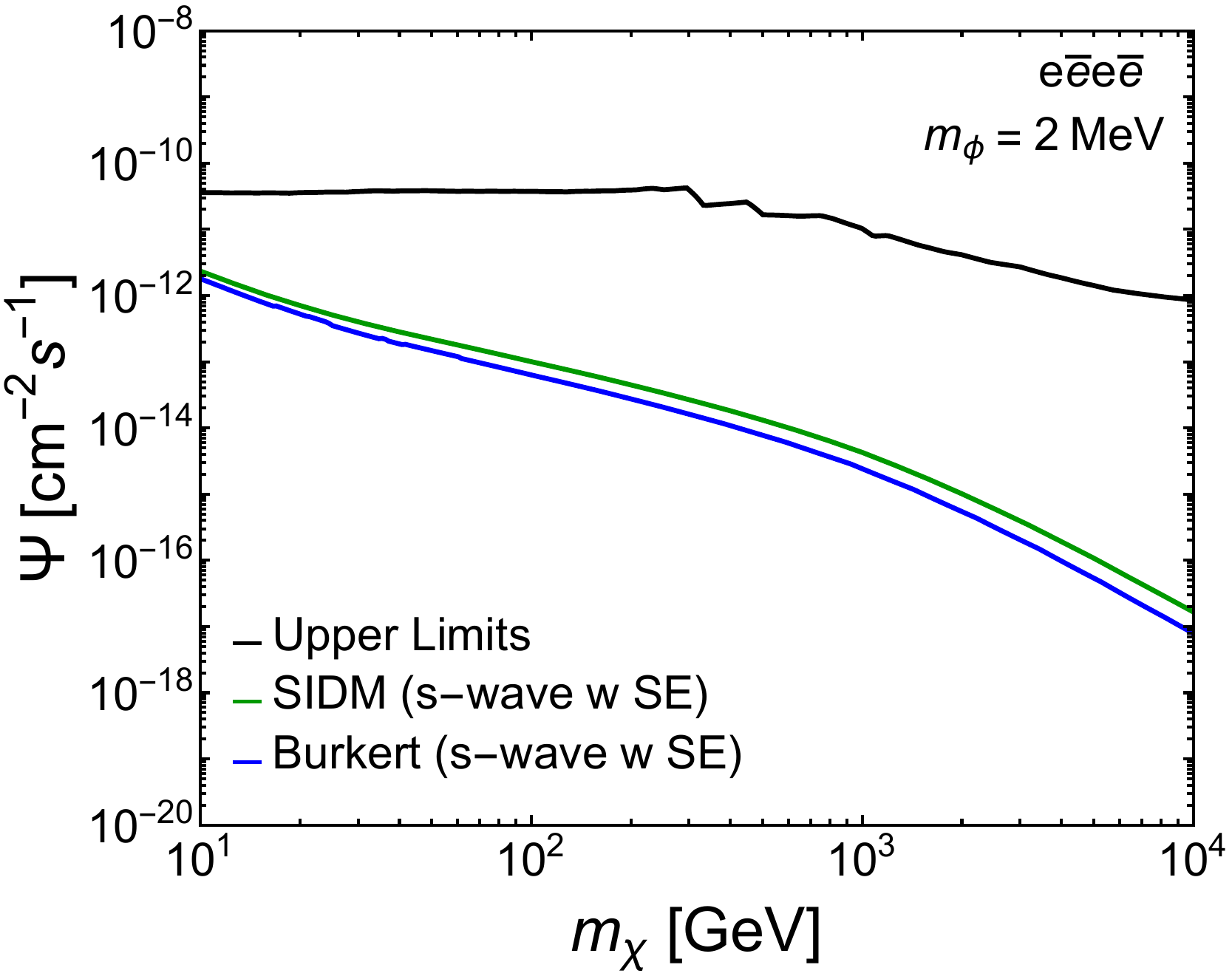}
    \caption{Left: Total mass profiles from the SIDM (green) and Burkert (blue) halo models, compared to the measurements~\cite{Oldham:2016,DeLaurentis:2022nrv} (data points with error bars). Middle: Dark matter density profiles for the SIDM (green) and Burkert (blue) models. Right: Corresponding projected gamma-ray fluxes for the s-wave $e\bar{e}e\bar{e}$ final states, including the Sommerfeld enhancement effect. The upper limits, derived from~\cite{Lacroix:2015lxa}, are shown for comparison (black). }
    \label{F:SIDM_DenistyMass}
\end{figure}

So far, we have adopted the cored Burkert density profile from~\cite{DeLaurentis:2022nrv} to model the M87 halo and have shown that constraints on dark matter annihilation from the spike surrounding the SMBH can be substantially weakened. In this section, we explore scenarios that could give rise to such a cored dark matter density profile. In CDM, active galactic nuclei (AGN) feedback associated with SMBHs can regulate galaxy formation and evolution~\cite{Fabian_2012}, potentially even altering dark matter distributions. For example, hydrodynamical simulations show that AGN feedback can reduce central densities within $\sim20~{\rm kpc}$ by a factor of $5\textup{--}10$ for CDM halos with a mass of $\sim8.8\times 10^{13}~{\rm M_\odot}$~\cite{Peirani:2016qvp}. However, these halos remain rather cuspy. In SIDM, interactions among dark matter particles can thermalize the inner halo and produce a large density core; see~\cite{Tulin:2017ara} for a review. Motivated by this, we perform an SIDM fit to the measured total mass profile of M87~\cite{Oldham:2015,Oldham:2016,DeLaurentis:2022nrv}, taking into account the effects of the baryonic potential on the SIDM halo profile~\cite{Kaplinghat:2013xca}.

We adopt a semi-analytical SIDM halo model~\cite{Kaplinghat:2013xca,Kaplinghat:2015aga} to derive the dark matter density profile for M87. This model assumes that the halo density profile follows an isothermal solution to the Jeans equation within a static baryonic potential in the inner regions, while retaining an NFW-like profile in the outer regions. The transition between these two regimes is defined by the radius at which dark matter self-scattering occurs once per Hubble time. For a given set of model parameters that characterize the initial CDM halo, the baryonic profile, and the self-scattering cross section, the model predicts the full density profile of the halo. Its accuracy has been validated against hydrodynamical SIDM simulations~\cite{Robertson:2017mgj,Robertson:2020pxj}. In practice, we use the publicly available code developed in~\cite{Jiang:2022aqw} to perform the numerical calculations.

We use a Hernquist profile~\cite{Hernquist:1990be} to model the stellar distribution of M87, which is more straightforward to implement in the semi-analytical SIDM halo model than the Nuker profile,
\begin{equation}
    \rho_\mathrm{H}=\frac{M_\mathrm{H}}{2\pi}\frac{r_\mathrm{H}}{r}\frac{1}{\left(r+r_\mathrm{H}\right)^3},
    \label{eq:hernquist}
\end{equation}
where $M_{\rm H}$ and $r_{\rm H}$ are the scale mass and radius, respectively. These parameters are determined by fitting Eq.~\ref{eq:hernquist} to the Nuker profile that characterizes the luminosity of M87. We fix the scale radius at $r_\mathrm{H}=4.65~\mathrm{kpc}$, as it reproduces the overall shape of the luminosity profile well, while $M_{\rm H}$ depends on the assumed mass-to-light ratio. If we adopt $M_{\rm sph}/L_\star=8.6~{\rm M_\odot L^{-1}_\odot}$ as in~\cite{DeLaurentis:2022nrv} (see the discussion in Sec.~\ref{sec:density}), the SIDM halo density in the inner regions becomes too high to be consistent with the total mass profile inferred for M87. This arises because an SIDM halo is more responsive to the baryonic potential than its CDM counterpart; with such a deep baryonic potential, the SIDM core shrinks and the density increases significantly compared to the SIDM-only case. We therefore relax the constraint on the mass-to-light ratio. After several trials, we find that $M_\mathrm{sph}/L_\star=3.5~{\rm M_\odot L^{-1}_\odot}$ yields a good fit, corresponding to $M_{\rm H} = 5.3\times 10^{11}~\mathrm{M_{\odot}}$. As a reference, Ref.~\cite{Oldham:2016} extensively studied the mass-to-light ratio of M87 by fitting four halo models, including both cuspy and cored profiles. They found $M_\mathrm{sph}/L_\star\approx6.9~{\rm M_\odot L^{-1}_\odot}$ assuming isotropic stellar velocity dispersion, and $M_\mathrm{sph}/L_\star\approx3.5~{\rm M_\odot L^{-1}_\odot}$ for anisotropic dispersion. This latter, anisotropic scenario is consistent with the expectation from our SIDM halo model for M87.

For the initial halo, we adopt a mass of $1.3 \times 10^{14}~\mathrm{M_{\odot}}$ and a concentration of $6$, close to the median value expected from the cosmological halo concentration-mass relation~\cite{Dutton:2014xda}. The self-scattering cross-section is fixed to be $\sigma/m =0.5 ~ \mathrm{cm^2 g^{-1}}$. For a viable self-interacting dark matter model, the self-scattering cross section must be velocity-dependent, decreasing to $\sigma/m \approx 0.1~{\rm cm^2 g^{-1}}$ for halos with masses of $10^{15}~{\rm M_\odot}$ to remain consistent with constraints from lensing measurements of galaxy clusters~\cite{Kaplinghat:2015aga,Sagunski:2020spe,Andrade:2020lqq}. In our analysis, we adopt a constant cross section of $\sigma/m = 0.5~{\rm cm^2~g^{-1}}$, which is broadly consistent with the effective cross section~\cite{Yang:2022hkm} in $10^{14}~{\rm M_\odot}$ halos predicted by recently proposed velocity-dependent SIDM models aimed at explaining the diversity of dark matter distributions~\cite{Nadler:2023nrd,Nadler:2025jwh}. We further set the age of the halo to be $10~{\rm Gyr}$ and compute the SIDM density profile using the public code developed in~\cite{Jiang:2022aqw}.

Fig.~\ref{F:SIDM_DenistyMass} (left panel) shows the total mass profiles from the SIDM halo model (green) and the Burkert halo model~\cite{DeLaurentis:2022nrv} (blue), compared to the measurements~\cite{Oldham:2016,DeLaurentis:2022nrv} (black dots with error bars). We see that our SIDM halo model fits the data well, comparably to the Burkert model. We have checked that the predicted total mass profile changes only mildly and remains within the measurement uncertainties when the mass-to-light ratio is varied in the range $M_\mathrm{sph}/L_*=3\textup{--}4~{\rm M_\odot L^{-1}_\odot}$. Fig.~\ref{F:SIDM_DenistyMass} (middle panel) shows the density profile of the SIDM halo (green), including the density spike. Compared to the Burkert profile (blue), the SIDM halo exhibits a higher density toward the inner regions due to the baryonic effects on the SIDM halo. Accordingly, the expected signal fluxes for the s-wave $e\bar{e}e\bar{e}$ final state are enhanced by a factor of $1.3\textup{--}2$ for $m_\chi=10~{\rm GeV}\textup{--}10~{\rm TeV}$, as shown in Fig.~\ref{F:SIDM_DenistyMass} (right panel), though they still lie well below the upper limits. Similar results are obtained for the other annihilation channels.

\section{Conclusions}
\label{sec:con}

We revisited the constraints on dark matter annihilation in M87. The SMBH at the center of M87 could induce a density spike, enhancing annihilation signals. Using a cored halo density profile for M87 based on recent kinematic measurements of multiple tracers, we constructed a spike profile for both SIDM and CDM models. We considered a dark matter model with a light mediator and calculated velocity-dependent $J$ and $Q$ factors, accounting for the Sommerfeld enhancement effect. For the cored Burkert density profile, the smooth halo component dominates the expected signal fluxes and the spike contribution is negligible. The projected gamma-ray fluxes are several orders of magnitude below observational upper limits. We further constructed an SIDM halo model for M87 that incorporates the effects of baryons on the dark matter distribution and found that the resulting projected fluxes are slightly enhanced but remain well below the observational upper limits. In the future, it would be interesting to extend this study by reexamining indirect detection constraints from other galaxies, such as the Milky Way. Additionally, one could directly fit semi-analytical SIDM halo models~\cite{Jiang:2022aqw,Yang:2023jwn,Yang:2024tba} to kinematic data from M87, as well as from other massive elliptical galaxies that favor a shallow dark matter density profile (see, e.g.,~\cite{Kong:2024zyw}). We leave these investigations for future work.

\acknowledgments
We thank Yi-Ming Zhong for useful discussion. This work was supported by the U.S. Department of Energy under Grant No. de-sc0008541.

\bibliography{references}

\end{document}